\documentclass[reprint,aps,pra,superscriptaddress,showpacs]{revtex4-1}
\usepackage{bm}
\usepackage{graphicx}
\usepackage{color}
\usepackage{amsmath}
\usepackage{natbib}
\usepackage{url}
\usepackage{textcase}
\usepackage{amsthm}
\usepackage{mathcomp}
\usepackage{amsfonts}
\usepackage{amssymb}
\usepackage{mathtools}
\usepackage{physics}

\begin{document}

\title{Chiral state-conversion without encircling an exceptional point}

\author{Absar U. Hassan}
\email{absar.hassan@knights.ucf.edu}
\affiliation{CREOL/College of Optics and Photonics, University of Central Florida, Orlando, Florida 32816, USA}
\author{Gisela L. Galmiche}
\affiliation{CREOL/College of Optics and Photonics, University of Central Florida, Orlando, Florida 32816, USA}
\author{Gal Harari}
\affiliation{Physics Department and Solid State Institute, Technion, Haifa 32000, Israel.}
\author{Patrick LiKamWa}
\affiliation{CREOL/College of Optics and Photonics, University of Central Florida, Orlando, Florida 32816, USA}
\author{Mercedeh Khajavikhan}
\affiliation{CREOL/College of Optics and Photonics, University of Central Florida, Orlando, Florida 32816, USA}
\author{Mordechai Segev}
\affiliation{Physics Department and Solid State Institute, Technion, Haifa 32000, Israel.}
\author{Demetrios N. Christodoulides}
\affiliation{CREOL/College of Optics and Photonics, University of Central Florida, Orlando, Florida 32816, USA}

\date{\today}

\begin{abstract}
Dynamically varying system parameters along a path enclosing an exceptional point is known to lead to chiral mode conversion. But is it necessary to include this non-Hermitian degeneracy inside the contour for this process to take place? We show that a slow enough variation of parameters, even away from the system's exceptional point, can also lead to a robust asymmetric state exchange. To study this process, we consider a prototypical two-level non-Hermitian Hamiltonian with a constant coupling between elements. Closed form solutions are obtained when the amplification/attenuation coefficients in this arrangement are varied in conjunction with the resonance detuning along a circular contour. Using asymptotic expansions, this input-independent mode conversion is theoretically proven to take place irrespective of whether the exceptional point is enclosed or not upon encirclement. Our results significantly broaden the range of parameter space required for the experimental realization of such chiral mode conversion processes.
\end{abstract}


\maketitle

Recent years have seen a surging interest in non-Hermitian systems - settings where the norm of a wavefunction is not conserved. Ideas emerging from this area have now pervaded various fields of physics and continue to attract significant research activity. One particular example along these lines was the introduction of a wide class of non-Hermitian Hamiltonians that respect parity-time (PT) symmetry~\cite{Bender1998,*BenderQM}. These PT-symmetric configurations can exhibit entirely real eigenvalue spectra in spite of the fact that the potentials involved are complex. Such complex arrangements can be readily implemented in the field of optics~\cite{Ruter,*Regensburger} since amplification and decay can be conveniently controlled in waveguide and cavity structures. A prominent feature of non-Hermitian settings is the existence of points of extreme degeneracy, better known as exceptional points (EPs). At EPs, not only do the eigenvalues of a system coalesce but so do the corresponding eigenvectors.

Lately, a number of intriguing processes have been observed in structures supporting EPs. These include for example, extreme sensitivity to perturbations~\cite{MercedehNature}, loss-induced transparency effects~\cite{Guo}, band merging~\cite{Bo} and unidirectional invisibility~\cite{ZinLin,*FengUnidirectional}, to name a few. The topological structure around these degeneracies is also richer in comparison with Hermitian degeneracies. A quasi-static encirclement of an EP allows the instantaneous eigenstates to swap with each other~\cite{Dembowski, Ostrovskaya} at the end of the parameter cycle while at the same time imparting a geometric phase~\cite{HeissPhysicsofEPs,Alexei-GeometricPhase}. On the other hand, if the same parameters are varied dynamically, i.e. when the field is forced to evolve alongside the parameters, only one of the eigenstates remarkably dominates at the end of the cycle~\cite{GWunner,Nimrod-ResonanceCoalescence,Atabek-VibrationalCooling}. Importantly, this effect occurs in a faithful manner and is inherently chiral. Experiments carried out in the microwave~\cite{Rotter-Nature} and optomechanical~\cite{Harris-Nature} domains have recently confirmed this unconventional behavior. In a recent paper~\cite{HassanEP}, an analytical explanation of this asymmetric (or chiral) mode selection mechanism was provided, along with the fact that this effect can persist even in the presence of nonlinearities. In addition, the possibility of a single-channel optical omni-polarizer was proposed. At this point, one might ask the following fundamental question: is this effect exclusive to contours that enclose an EP or is it more generic in nature?

Here we analytically show that this phenomenon can manifest itself under more general conditions. In other words, chiral state conversion can faithfully occur even when the trajectory along which the parameters are varied does not enclose an exceptional point. We demonstrate that in the limit of slow variations, only a single eigenstate will emerge at the output, regardless of initial conditions. In a two-level system, the sense of rotation of the parameters, clock-wise (c.w.) or counter-clockwise (c.c.w.) is the only factor that determines the prevailing eigenstate. The model considered here is exactly solved in terms of Bessel functions, thus allowing one to continuously track the modal populations. Asymptotic expansions of these functions furnish an analytical proof of chiral mode selection even for non-EP enclosing loops.
To analyze the dynamics of two coupled states, constantly exchanging energy in space or time, we consider a non-Hermitian Hamiltonian undergoing circular variations in the diagonal terms. For example, in optics, this can be implemented using dielectric cavities (in the temporal $t$-domain) or waveguides (in the spatial $z$-coordinate). For this $2\times2$ system, the modal dynamics obey a Schr\"{o}dinger type equation,

\begin{equation}
 \label{full system1}
 i\frac{d}{dt}\ket{\Psi(t)}=
\begin{pmatrix}
ig(t) + \delta(t)   & -1       \\
-1           & -ig(t) - \delta(t)
\end{pmatrix}\ket{\Psi(t)}.
\end{equation}

In Eq.~(\ref{full system1}), the state vector $\ket{\Psi(t)}=[a(t),b(t)]^T$ describes the fields in the two coupled elements. The quantities $g(t)$ and $\delta(t)$ represent variations in the gain/loss and resonance-detuning, respectively. All variables are normalized with respect to the coupling constant (off-diagonal entries). Based on the values $g(t)$ and $\delta(t)$ assume at any given time $t$, the instantaneous eigenvalues $\lambda_i$ and eigenvectors $\ket{\psi_i}$ can always be found by using the ansatz $\ket{\Psi}=\ket{\psi_i} e^{-i\lambda_i t}$. A circular trajectory in the $(g,\delta)$-space can be parametrically described by: $g(t)=g_0-\rho\cos(\gamma t)$, $\delta(t)=\rho\sin(\gamma t)$. This circle, centered at $g_0$, has a radius of $\rho$ while $\gamma$ provides a measure as to how slowly the variations are performed. A c.w. (c.c.w.) loop is described by $\gamma>0$ ($\gamma<0$). In this framework, $g$ and $\delta$ return to their initial values at the end ($\tilde{T}=2\pi\gamma^{-1}$) of a cycle. Hence the forthcoming discussion is centered on the two eigenvectors corresponding to the terminal point $(g,\delta)=(g_0-\rho,0)$. At this location, one obtains,
\begin{equation}\label{eigenvectors}
  \ket{\psi_{1}}=\begin{bmatrix}1\\e^{i\theta}\end{bmatrix},
  \ket{\psi_{2}}=\begin{bmatrix}1\\-e^{-i\theta}\end{bmatrix}.
\end{equation}
The corresponding eigenvalues are given by, $\lambda_{1,2}=\pm\cos\theta$, where $\theta$ can be obtained from $\sin\theta=(g_0-\rho)$. The two vectors in Eq.~(\ref{eigenvectors}) are biorthogonal with the left eigenvectors $\ket{\tilde{\psi}_{1,2}}=[1,\pm e^{\mp i\theta}]^T$. In this non-Hermitian arrangement, the point of extreme degeneracy (EP) occurs in parameter space at $g=1$ and $\delta=0$, where the two eigenvalues coalesce at $\lambda_{1,2}=0$ and the corresponding eigenvectors collapse to $\ket{\psi_{1,2}}=[1,i]^T$. As previously shown, if this EP is enclosed in a single cycle parameter loop, then any input state will robustly transform into $\ket{\psi_{1}}$ if the encirclement is carried out in a c.w. sense. However, upon changing the direction of encirclement to c.c.w., $\ket{\psi_{2}}$ instead dominates the output~\cite{Uzdin-Ontheobservability}.

In what follows, we will reexamine this same effect even when the EP is excluded from the parameter loop. To do so, we recast Eq.~(\ref{full system1}) into a second order differential equation for the first element $a(t)$, according to,
\begin{equation}
 \label{2ndOrder-a}
\frac{d^2a(t)}{dt^2}=\left[ \rho^2e^{2i\gamma t} - \rho(2+i\gamma)e^{i\gamma t} + g_0^2 - 1 \right]a(t).
\end{equation}

A similar equation also holds for $b(t)$. For all practical purposes, it is sufficient to solve the system of Eq.~(\ref{full system1}) [and Eq.(\ref{2ndOrder-a})] for a c.w. trajectory only. Solutions obtained as such can then be directly used to describe the counterpart c.c.w. case, simply by employing the transformation $(a,b)\rightarrow(a^*,-b^*)$ or by replacing $\gamma$ with $-\gamma$, see supplementary of Ref.~\cite{HassanEP}. In general, Eq.~(\ref{2ndOrder-a}) can be solved through hypergeometric functions~\cite{HassanEP}. However, considerable insight into the system's behavior can be gained if we assume small encirclements, $\rho\ll1$, which allows the $\rho^2$ term to be neglected in Eq.~(\ref{2ndOrder-a}). In this case, this system can be reduced to a modified Bessel differential equation of $\nu^{\text{th}}$ order. For example, by employing the substitution $x=x_0e^{i\gamma t/2}$ where $x_0=2\gamma^{-1}\sqrt{\rho(2g_0+i\gamma)}$, one finds that $\nu=2\gamma^{-1}\sqrt{1-g_0^2}$. As in previous studies, $g_0=1$ can be used to describe contours that are centered at the EP. Here on the other hand, we allow $g_0$ to vary in the domain $[0,1]$ when $\rho$ is sufficiently small. Under these conditions, one finds that $a(t)=c_1I_{\nu}(x_0e^{i\gamma t/2})+c_2K_{\nu}(x_0e^{i\gamma t/2})$ where $c_{1,2}$ depend on initial conditions and $I_{\nu},K_{\nu}$ are Bessel functions of the first and second kind respectively, of order $\nu$. From here, one can directly determine the field $b(t)$ in the second element from Eq.~(\ref{full system1}). Therefore, the complete solution of this problem is given by,
\begin{widetext}
\begin{subequations}\label{CompleteSolution}
\begin{gather}
\ket{\Psi(t)}=\begin{bmatrix} I_{\nu}(x_0e^{i\frac{\gamma t}{2}})                              &                          K_{\nu}(x_0e^{i\frac{\gamma t}{2}}) \\
    i(g_0-\rho e^{i\gamma t})I_{\nu}(x_0e^{i\frac{\gamma t}{2}})-i\frac{dI_{\nu}(x_0e^{i\gamma t/2})}{dt}  &   i(g_0-\rho e^{i\gamma t})K_{\nu}(x_0e^{i\frac{\gamma t}{2}})-i\frac{dK_{\nu}(x_0e^{i\gamma t/2})}{dt}       \end{bmatrix}
    \begin{bmatrix} c_1\\c_2 \end{bmatrix}\\
    \begin{bmatrix} c_1\\c_2 \end{bmatrix}=-\frac{2}{\gamma}
    \begin{bmatrix} i(g_0-\rho)K_{\nu}(x_0)-i\frac{dK_{\nu}(x_0)}{dt}                 &                          -K_{\nu}(x_0)                        \\
                    -i(g_0-\rho)I_{\nu}(x_0)+i\frac{dI_{\nu}(x_0)}{dt}                &                           I_{\nu}(x_0)             \end{bmatrix}
    \ket{\Psi(0)}.
\end{gather}
\end{subequations}
\end{widetext}
\begin{figure*}
  \includegraphics{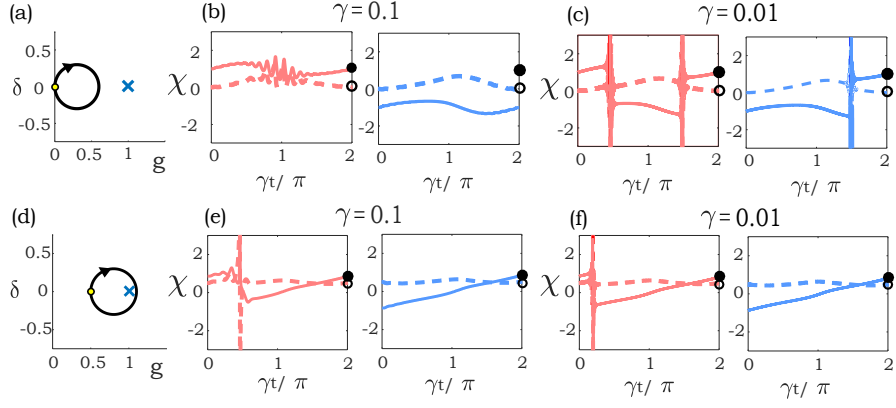}
  \caption{\label{Fig1}
  Two different c.w. parameter cycles are shown in (a) and (d) along with the ensuing behavior of $\chi$ in the corresponding panels [(b),(c)] and [(e),(f)] in each row. The loop in (a) lies far away from the EP (EP is shown as a cross)\textemdash with $(g_0,\rho)=(0.3,0.3)$. In the one shown in (d), the contour includes the EP\textemdash with $(g_0,\rho)=(0.8,0.3)$. The terminal points, where the two eigenvectors $\ket{\psi_{1,2}}$ are found, are marked by a yellow circle and the arrow shows the direction of encirclement. In (b) and (e), the resulting variation in $\chi$ at all times is shown when the rate of cycling is relatively large, i.e. $\gamma=0.1$. Plots on the left (shown in red) depict the case when the system is excited with $\ket{\psi_1}$ and those on the right (shown in blue) provide results for excitations with $\ket{\psi_2}$. In these plots, solid (dashed) lines represent real (imaginary) parts of $\chi$. As mentioned in the text, for this c.w. cycle, the state expected at the output is $\ket{\psi_1}$ that corresponds to $\chi\rightarrow e^{i\theta}$. The real (imaginary) part of this expected result is shown as a filled (empty) circle at $t=2\pi\gamma^{-1}$. In (c) and (f), the rate of cycling is reduced to $\gamma=0.01$ and both excitations end up at the correct location even for the non-EP enclosing case\textemdash (c). Although mode conversion is not robust in (b)\textemdash consider the plot on the right, results for the EP-inclusive loop show robust state conversion not only when the encirclement is slow [in (f)], but also when it is fast $\gamma=0.1$, as in (e).}
\end{figure*}
Equations~(\ref{CompleteSolution}) describe the behavior of the fields in both elements at all times, from the start to the end of the cycle. For slow cycles where $\gamma\ll1$, the Bessel functions can be asymptotically expanded in terms of elementary functions in order to find the modal content at the output in terms of the eigenvectors $\ket{\psi_{1,2}}$. To do so, notice that the time variable only appears in the form of an exponential and results in a complex phase inside the arguments of the Bessel functions. In this regard, the end of the parameter loop amounts to acquiring a phase of $e^{i\pi}$ which reduces the analysis to finding expansions of: $I_{\nu}(x_0e^{i\pi}),K_{\nu}(x_0e^{i\pi}),J_{\nu}'(x_0e^{i\pi})\text{ and }Y_{\nu}'(x_0e^{i\pi})$, where prime indicates a time derivative. Expansions for either a large argument or a large order do not apply since both $\nu$ and $x_0$ happen to be large (even though $\nu>|x_0|$) as $\gamma\rightarrow0$. In fact, one requires uniform asymptotic expansions as given in Eqs.(10.41.3-6) of Ref.~\cite{DLMF} where the order and argument simultaneously go to large values. Moreover, to accommodate (cancel out) the extra phase of $e^{i\pi}$, we use analytic continuation according to,
\begin{subequations}\label{AnalyticContinuation}
\begin{gather}
I_{\nu}(x_0e^{i\frac{\gamma t}{2}})=e^{i\nu\pi}I_{\nu}(x_0e^{i\frac{\gamma t}{2}}e^{-i\pi})\\
K_{\nu}(x_0e^{i\frac{\gamma t}{2}})=e^{-i\nu\pi}K_{\nu}(x_0e^{i\frac{\gamma t}{2}}e^{-i\pi})-i\pi I_{\nu}(x_0e^{i\frac{\gamma t}{2}}e^{-i\pi}).
\end{gather}
\end{subequations}
Finally, to conform with the notation in Ref.~\cite{DLMF}, we substitute $z=x_0/\nu$, $\eta=\sqrt{1+z^2}+\ln[1+\frac{z}{1+\sqrt{1+z^2}}]$, and using the fact that $e^{\nu\eta}\gg e^{-\nu\eta}$, the modal fields at the end of the parameter cycle ($\tilde{T}=2\pi\gamma^{-1}$) assume a simplified form,
\begin{widetext}
\begin{equation}\label{TendExpansion}
\ket{\Psi(\tilde{T})}\sim \frac{e^{i\nu\pi}}{\sqrt{2\nu\pi}}(1+z^2)^{-\frac{1}{4}}e^{\nu\eta}\begin{bmatrix}
1     &         -i\pi e^{-i\nu\pi} \\
i(g_0-\rho)+\frac{\gamma x_0}{2}\sqrt{1+\frac{1}{z^2}}     &    -i\pi e^{-i\nu\pi}\left[i(g_0-\rho)+\frac{\gamma x_0}{2}\sqrt{1+\frac{1}{z^2}}\right]     \end{bmatrix}
\begin{bmatrix} c_1\\c_2 \end{bmatrix}.
\end{equation}
\end{widetext}
To see whether $\ket{\Psi(\tilde{T})}$ has any semblance of $\ket{\psi_{1,2}}$, it is more convenient to convert it into the form $\ket{\Psi(\tilde{T})}\propto[1,\chi]^\text{T}$ where $\chi=b(\tilde{T})/a(\tilde{T})$. Mode conversion into $\ket{\psi_1}$ or $\ket{\psi_2}$ would be established only if $\chi\rightarrow e^{i\theta}$ or $\chi\rightarrow -e^{-i\theta}$, respectively. From Eq.~(\ref{TendExpansion}) it can be found that $\chi\rightarrow i(g_0-\rho)+\frac{\gamma x_0}{2}\sqrt{1+z^{-2}}$, and substituting back $z=x_0/\nu$ we directly find that,
\begin{equation}\label{Chi_end_cw}
  \chi_{\text{c.w.}}\rightarrow i(g_0-\rho)+\sqrt{1-g_0^2+2\rho g_0}=e^{i\theta}.
\end{equation}
Thus the final state at the output is indeed only $\ket{\psi_1}$. The result in Eq.~(\ref{Chi_end_cw}) holds for a c.w. loop since $\gamma>0$ was assumed. If instead, one solves for the c.c.w. loop, the only change in the analysis will be a reflection $\gamma\rightarrow-\gamma$. In this regard, the following relation can be immediately inferred,
\begin{equation}\label{Chi_end_ccw}
  \chi_{\text{c.c.w.}}\rightarrow i(g_0-\rho)-\sqrt{1-g_0^2+2\rho g_0}=-e^{-i\theta}.
\end{equation}
Here, mode conversion occurs into the state $\ket{\psi_2}$ instead. The above results not only prove that robust state conversion is possible even in the absence of an EP-encirclement, but also reaffirm the inherent chirality of this process. The reason why this mechanism is input-independent becomes clear after inspecting the matrix elements of $M$ in Eq.~(\ref{TendExpansion}) when it is written as $\ket{\Psi(\tilde{T})}\sim M[c_1,c_2]^T$. For example, in the c.w. case, both $m_{21}=e^{i\theta}m_{11}$ and $m_{22}=e^{i\theta}m_{12}$, which directly leads to the conclusion that $\chi$ is independent of any initial conditions $a(0)$,$b(0)$. Similar conclusions hold for a c.c.w. contour.

At this point, one might ask what benefit, if any, accrues from including the EP in a dynamic parameter contour, if mode conversion can take place even through non-EP encircling loops? To this end, we found that for a given rate of change of parameters (or adiabaticity $\gamma$), mode conversion becomes more robust as the contour moves closer to and eventually encloses the EP. However, in the limit of very slow cycles ($\gamma\ll1$), enclosing or not enclosing an EP in a parameter loop makes no significant difference. In Fig.~\ref{Fig1}, we show two distinct c.w. circular trajectories on the $(g,\delta)$ plane [(a) and (d)], with relatively large values of the radius $\rho$, along with the ensuing behavior of $\chi$ when $\gamma=0.1$ in the middle column [(b) and (e)] and for $\gamma=0.01$ in the last column [(c) and (f)]. In doing so, the fields were expressed as $\propto[1,\chi]^{\text{T}}$ at all times. The c.c.w. case has similar behavior only this time $\chi\rightarrow-e^{-i\theta}$ instead. When the cycle is not adiabatic enough, both initial eigenvectors $\ket{\psi_{1,2}}$ do not end up in the anticipated location ($\chi\rightarrow e^{i\theta}$), see the right-plot in (b). However, when $\gamma$ is reduced to $\gamma=0.01$ (more adiabatic), state conversion is apparent in (c), even in the non-EP inclusive case. On the other hand, when the EP is located inside the parameter path (lower panel), robust mode conversion takes place for both large (e) and small (f) values of adiabaticity $\gamma$. We would also like to point out that for paths that are not circular, e.g. deformed loops, one can still obtain the aforementioned chiral mode conversion\textemdash another indication as to how robust this mechanism is.

In conclusion, we have theoretically demonstrated that asymmetric (or chiral) mode conversion is not exclusive to parameter loops that include an EP. Instead, it can take place even if the contour lies in the vicinity of an EP, provided that the parameters are varied in an adiabatic manner. An immediate ramification of this result could be the potential reduction of the amount of gain-loss contrast needed in coupled optical configurations to observe this effect. For example, instead of cycling around $g=\kappa$ (where $\kappa$ is the coupling strength), one can achieve the same outcome by staying longer around, say $g=0.1\kappa$\textemdash hence reducing the range of gain needed by an order of magnitude. This could facilitate the observation of the aforementioned chiral effects in metasurface and optical waveguide arrangements~\cite{Yidong-ChiralEPs,*Yidong-SciRep}, as well as in atomic systems~\cite{GWunnerPRA}. The inclusion of nonlinearities in such non-EP enclosing loops could be an interesting aspect for further study. It would also be worthwhile to investigate the sudden transitions occurring during the cycle of such a non-Hermitian evolution in terms of the Stokes phenomenon of asymptotics~\cite{Uzdin-Berry} and stability loss effects~\cite{Rotter-PRA}.



\bibliography{References}

\end{document}